\documentclass{article}
\usepackage{spconf,amsmath,amssymb,graphicx,booktabs,hyperref}
\usepackage{multirow}
\usepackage{tikz}
\usetikzlibrary{arrows.meta,positioning,shapes.misc,shapes.geometric,fit,calc}

\title{NEONATAL EEG SEIZURE PREDICTION WITH QUANTUM-INSPIRED CONTINUOUS-VARIABLE (BOSONIC) RESERVOIR COMPUTING}

\name{
  Krishna Bhatia$^{1}$,
  Chun-Yu Lin$^{2}$
  \thanks{QuantumAI Lab, Fractal AI Research, Fractal Analytics, India}
}
\address{
$^{1}$QuantumAI Lab, Fractal AI Research, Fractal Analytics, Mumbai, India \\
{\tt krishna.bhatia@fractal.ai} \\
$^{2}$National Center for High-performance Computing, National Institutes of Applied Research, Hsinchu, Taiwan \\
{\tt lincy@nchc.org.tw}
}

\begin{document}
\maketitle

\begin{abstract}
We study short-horizon neonatal seizure prediction on Helsinki-style (HUH) preprocessed EEG feature tables using quantum-inspired reservoir computing. Under strict subject-wise cross-validation on imbalanced preictal--interictal windows, we test whether reservoir constructions shaped after (i) gate-based qubit dynamics and (ii) continuous-variable (bosonic/photonic) dynamics provide a practical sequence encoder relative to strong classical baselines. Both reservoirs are executed here purely in classical simulation on a single GPU; accordingly, we do not claim quantum advantage, but rather evaluate whether quantum-shaped dynamics act as useful, low-footprint random feature maps with a trained linear readout. We compare against Random Forests and RBF--SVMs, and report AUROC, AUPRC, per-subject AUPRC, expected calibration error, and stratified bootstrap confidence intervals. Across our experiments, the CV/bosonic reservoir is consistently competitive and can improve AUPRC over the baselines, while the gate-style reservoir is more sensitive to hyperparameters. These results suggest that quantum-inspired, photonics-motivated reservoirs can be practical sequence encoders for neonatal EEG feature tables under conservative subject-wise evaluation.
\end{abstract}

\section{Introduction}
Neonatal seizures are early markers of brain injury but are often clinically silent, so continuous multichannel EEG is used in the NICU. Because expert review is scarce, automated pipelines based on hand-crafted spectral, amplitude, asymmetry, and connectivity features followed by SVM-like classifiers have been widely studied~\cite{Temko2011EEGAmplitudeFreq,Temko2011SVMNeonatal}.
 Later, CNN and CNN--RNN models on EEG-derived time--frequency images improved sensitivity and reduced false positives~\cite{Oshea2017NeonatalCNN,Oshea2018NeonatalCNN}, but these models are heavier, need more data, and can struggle with the missingness and per-subject heterogeneity typical of clinical EEG tables.

More recent neonatal work explores prediction and large-scale detection with deep learning on quantitative or raw EEG~\cite{Kim2025ShortHorizonNeonatal,Hogan2025ScalingNeonatalCNN}; in contrast, we focus on the clinically common feature-table setting and ask whether quantum-shaped reservoirs provide a lightweight sequence embedding under strict subject-wise splits.

In parallel, quantum machine learning (QML) proposed to use quantum evolutions as fixed, high-dimensional feature maps with only a classical readout~\cite{Schuld2015IntroQML}. Quantum reservoir computing (QRC) is a convenient instance: a randomly parametrized quantum system is driven by an input sequence and only a simple output layer is learned~\cite{FujiiNakajima2017Harness,Nakajima2020QRC}. Continuous-variable and photonic variants show how classical signals can be injected as small displacements, mixed by random interferometers, and weakly nonlinearized~\cite{Killoran2019CVQNN,MartinezPena2021IPC}.

We take a realistic, Helsinki-style (HUH) neonatal EEG feature table (subjects, absolute time indices, seizure/preictal flags, numeric descriptors), build per-subject fixed-length sequences (18 steps, $\approx$6 minutes), and test whether such \emph{quantum-shaped} reservoirs provide measurable gains over strong classical baselines under subject-wise cross-validation on imbalanced preictal--interictal labels. The pipeline standardizes on train folds only, imputes clinical NaNs, and reports AUROC, AUPRC, bootstrap uncertainty, calibration, and per-subject AUPRC. We use two encoders: (i) a gate-style, stateful reservoir that unrolls the whole sequence in one circuit-like evolution and reads out single- and two-body observables, and (ii) a continuous-variable (bosonic) reservoir that projects features to multiple modes, applies randomized linear mixing with weak nonlinearity, and time-multiplexes to expand dimensionality. Both run fully quantum-inspired on GPU. In this setup, the CV reservoir is the most stable and often matches or slightly exceeds tree and kernel baselines, while the gate-style reservoir is competitive but more hyperparameter-sensitive.

\section{Related Work}
Classical feature-based neonatal EEG seizure detection has been shown to work in clinical conditions and to tolerate strong class imbalance~\cite{Temko2011ClinicalNeonatal}. Deep learning models operating on raw EEG or time--frequency representations (CNNs and variants) have improved neonatal seizure \emph{detection} when large annotated corpora are available~\cite{Oshea2017NeonatalCNN,Oshea2018NeonatalCNN,Hogan2025ScalingNeonatalCNN}. For \emph{short-horizon prediction}, recent work demonstrated that minute-scale neonatal seizure risk estimation is feasible with deep learning on quantitative EEG~\cite{Kim2025ShortHorizonNeonatal}, though external validation and careful split protocols remain important.

Independently, quantum reservoir computing (QRC) showed that driving a fixed, randomly parametrized quantum dynamical system and training only a linear readout can yield powerful sequence processors~\cite{FujiiNakajima2017Harness,Nakajima2020QRC}. Martínez-Pe\~na \emph{et al.} analyzed the information-processing capacity of reservoirs and emphasized the ``large linear memory with weak nonlinearity'' regime~\cite{MartinezPena2021IPC}. On the continuous-variable side, photonic primitives---displacements, interferometers, and weak nonlinearities---provide a natural recipe for CV reservoirs and CV quantum neural networks~\cite{Killoran2019CVQNN}. Importantly, the broader reservoir-computing community has also demonstrated fast, hardware-friendly photonic reservoirs that motivate CV-style designs~\cite{Aadhi2025ScalablePhotonicRC}, and recent proposals outline feasible photonic quantum reservoirs using photon-number-resolving measurements~\cite{Nerenberg2025PNR_QRC}.

Our work connects these strands by keeping the clinically realistic feature-table setting and strict subject-wise evaluation, while swapping the sequence encoder for (i) a gate-style and (ii) a bosonic/CV-style \emph{quantum-inspired} reservoir that runs fully in software. The goal is not to claim quantum advantage, but to test whether \emph{quantum-shaped} dynamics provide a practical low-footprint embedding in a conservative neonatal generalization regime.

\section{Problem Definition and Model}
\begin{figure}
    \centering
    \includegraphics[width=\linewidth]{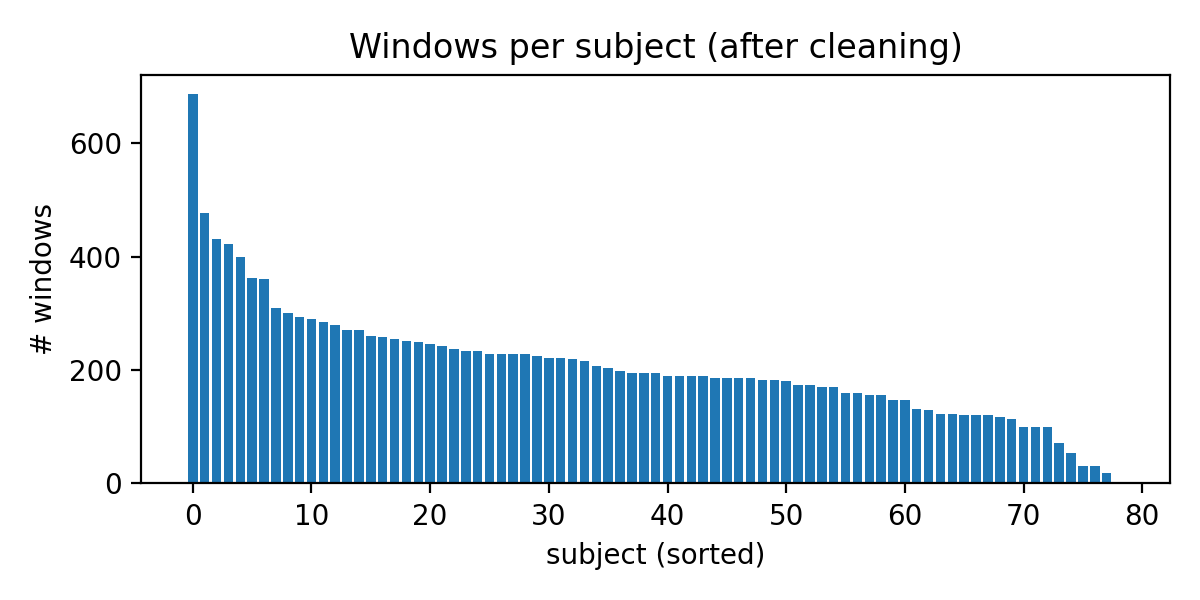}
    \caption{Per-subject sliding windows ($L{=}18$) after cleaning.}
    \label{fig:datawindows}
\end{figure}

Let the cleaned, sequentialized dataset be
\[
\mathcal{D} = \{(\mathbf{X}_i, y_i, g_i)\}_{i=1}^N,
\]
where $g_i$ is the subject identifier (\texttt{subject\_id}), $y_i \in \{0,1\}$ is the preictal label of the \emph{last} time step in the window (after discarding seizing/invalid rows), and $\mathbf{X}_i \in \mathbb{R}^{L \times D}$ is a length-$L$ sequence of $D$ numeric EEG features, ordered by absolute time index ($L=18$ in our experiments). Windows are built separately for each subject; subjects with fewer than $L$ valid rows are skipped. We construct a subject-wise partition
\(\{(\mathcal{T}_k, \mathcal{V}_k)\}_{k=1}^K\) with $K=5$ using GroupKFold so that no subject appears in both train and validation, matching the goal of generalizing to unseen neonates. The dataset windows per subject are demonstrated in~\autoref{fig:datawindows}

\subsection{Stateful gate-style reservoir}
We instantiate a fixed, random circuit template of depth $d$ over $Q$ logical units. A subset $\mathcal{S} \subseteq \{1,\dots,Q\}$ of size $|\mathcal{S}| = \lfloor \rho Q \rfloor$ is chosen to receive inputs. 
Here $\rho \in (0,1]$ is the \emph{injection fraction}, i.e.,
\[
\rho \approx \frac{|\mathcal{S}|}{Q},
\]
so larger $\rho$ means more units directly receive the input at each step. For a sequence $\mathbf{X} \in \mathbb{R}^{L \times D}$ we first project each step through a random matrix $P \in \mathbb{R}^{|\mathcal{S}|\times D}$ to obtain
\begin{equation}
A_t = P \, \mathbf{X}_t \in \mathbb{R}^{|\mathcal{S}|}, \qquad t = 1,\dots,L.
\end{equation}
The reservoir evolution is then
\begin{equation}
U(\mathbf{X}) = \prod_{t=1}^{L} \bigl( U_{\mathrm{ent}} \circ U_{\mathrm{inj}}(A_t) \bigr),
\end{equation}
where $U_{\mathrm{inj}}$ maps the projected features into rotations on the injection set and $U_{\mathrm{ent}}$ is a fixed mixing block (e.g.\ nearest-neighbour interactions plus local phases). After processing the window we use the final state (optionally with a few pairwise products) as the fixed feature map:
\begin{equation}
\boldsymbol{\phi}_{\mathrm{gate}}(\mathbf{X})
=
\bigl[\mathbf{r}_L^\top,\; \{r_{L,a}r_{L,b}\}_{(a,b)\in\mathcal{P}}\bigr].
\end{equation}

\subsection{Continuous-variable (bosonic) reservoir}
\label{ssec:cvrc}
We model a CV-style reservoir (displacement injection, fixed mixing, weak nonlinearity, time multiplexing) in classical simulation.

\subsubsection{Input projection.}
Given a sequence $\mathbf{X} \in \mathbb{R}^{L \times D}$, we operate on $M$ modes ($M \ll D$) and fix a random projection
\begin{equation}
\mathbf{P} \in \mathbb{R}^{M \times D}, \qquad P_{ij} \sim \mathcal{N}\!\bigl(0,\sigma_P^2\bigr), \qquad \sigma_P=\frac{1}{\sqrt{D}},
\end{equation}

so that each time step $\mathbf{x}_t$ is mapped to a displacement vector
\begin{equation}
\boldsymbol{\alpha}_t = \mathbf{P}\,\mathbf{x}_t \in \mathbb{R}^{M}.
\end{equation}

\subsubsection{Displacement and mixing.}
We represent the reservoir state as $\mathbf{s}_t = [\mathbf{q}_t^\top, \mathbf{p}_t^\top]^\top \in \mathbb{R}^{2M}$ and inject data by
\begin{equation}
\tilde{\mathbf{q}}_t = \mathbf{q}_{t-1} + \gamma \boldsymbol{\alpha}_t,\qquad
\tilde{\mathbf{p}}_t = \mathbf{p}_{t-1},
\end{equation}
with small $\gamma>0$. A pre-sampled symplectic-like matrix $\mathbf{S} \in \mathbb{R}^{2M \times 2M}$, built once per run, performs the interferometric mixing
\begin{equation}
\hat{\mathbf{s}}_t = \mathbf{S}
\begin{bmatrix}
\tilde{\mathbf{q}}_t \\ \tilde{\mathbf{p}}_t
\end{bmatrix},
\end{equation}
so injected information is spread across modes and quadratures.

\subsubsection{Weak nonlinearity.}
To avoid a purely linear predictor we apply a pointwise, bounded nonlinearity that mimics weak Kerr effects:
\begin{equation}
\mathbf{s}_t = \tanh\!\bigl(\hat{\mathbf{s}}_t + \eta \hat{\mathbf{s}}_t^{\odot 3}\bigr),
\end{equation}
with small $\eta$. This follows the “small nonlinearity, large linear memory” recipe in quantum-RC capacity studies~\cite{MartinezPena2021IPC}.

\subsubsection{Time multiplexing and readout.}
For each external step $t$ we repeat the injection–mixing nonlinearity chain $V$ times with the same $\boldsymbol{\alpha}_t$, yielding intermediate states $\mathbf{s}_t^{(1)},\dots,\mathbf{s}_t^{(V)}$. We emit the $q$-parts,
\begin{equation}
\mathbf{z}_t = \bigl[\mathbf{q}_t^{(1)\top}, \dots, \mathbf{q}_t^{(V)\top}\bigr]^\top \in \mathbb{R}^{VM},
\end{equation}
and concatenate over $t$ to obtain
\begin{equation}
\boldsymbol{\phi}_{\text{cv}}(\mathbf{X}) = [\mathbf{z}_1^\top, \dots, \mathbf{z}_L^\top]^\top \in \mathbb{R}^{LVM}.
\end{equation}
A class-balanced ridge classifier
\begin{equation}
f_{\text{cv}}(\mathbf{X}) = \mathbf{w}^\top \boldsymbol{\phi}_{\text{cv}}(\mathbf{X}) + b
\end{equation}
is then trained per fold. In practice, $M\in\{24,28,32\}$ and $V\in\{6,8\}$ were sufficient to reach the strongest validation scores.

\subsection{Uncertainty and calibration}
\label{ssec:uncal}
For every model and fold we compute AUROC and AUPRC on the held-out subjects and estimate uncertainty with stratified bootstrap ($B = 1000$)~\cite{Efron1979Bootstrap} over validation predictions,
\begin{equation}
\hat{\theta}^{(b)} = \mathrm{AUPRC}(\mathcal{V}_k^{(b)}), \qquad b=1,\dots,B,
\end{equation}
reporting the mean and the $2.5$/$97.5$ percentiles. We also compute expected calibration error (15 bins)~\cite{Guo2017Calibration} on sigmoid-transformed scores and inspect per-subject AUPRC to capture heterogeneity of difficulty across neonates.

\section{Experimental Setup}
\begin{figure}[h]
    \centering
    \includegraphics[width=\linewidth]{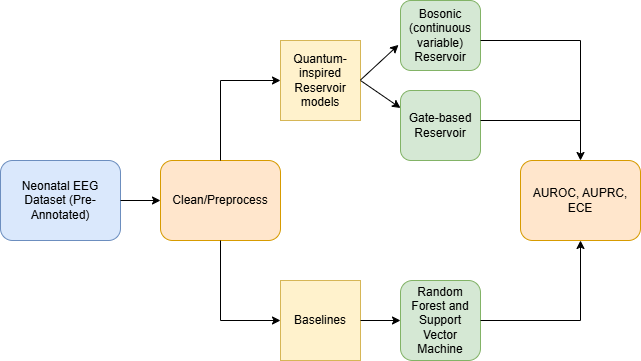}
    \caption{Pipeline overview: preprocessing, reservoirs, and readout.}
    \label{fig:flowchart}
\end{figure}

We work on the Helsinki-style (HUH) neonatal EEG CSVs provided with the dataset~\cite{Stevenson2019NeonatalDataset}: they contain subject identifiers, an absolute time index, booleans for seizure and channel validity, and a large number of numeric EEG features. We remove rows marked as seizing, keep only valid rows, drop explicit augmentations, and build sliding windows of $L=18$ steps per subject. All preprocessing (standardization, NaN imputation) is fitted on train folds only and applied to validation folds to prevent information leakage. Classical baselines are: (i) Random Forest with 600 trees, class-weighted, (ii) RBF--SVM with a regularization parameter $C=10$, a kernel coefficient set by scikit-learn's \texttt{scale} heuristic (inverse of the feature count times the data variance, i.e.\ $\gamma=\texttt{scale}$), and probability estimates enabled. We also included a small mixed reservoir (ResMix) that concatenates gate and CV features and fits a single ridge readout (Fig. \ref{fig:flowchart}).

\subsection{Sensitivity and compute}
Both reservoirs are randomized feature maps; performance depends on effective state dimension and avoiding $\tanh$ saturation (``memory + weak nonlinearity'')~\cite{MartinezPena2021IPC}. In targeted one-factor ablations around stable defaults, moderate $M$ and $V$ were sufficient for the CV reservoir, while overly large $\gamma$ or $\eta$ tended to degrade calibration; the gate-style variant was more sensitive to depth/size without rebalancing injection, consistent with prior QRC observations~\cite{FujiiNakajima2017Harness}. The CV readout has $LVM$ weights (e.g., $2592$ for $L{=}18,M{=}24,V{=}6$) and inference is dominated by $LV$ mixing steps, making it lightweight and batchable on GPU.

\section{Results}
Table~\ref{tab:results} reports the strongest subject-wise validation performance observed among the configurations we explored for each model family under the same 5-fold GroupKFold protocol. Because the dataset is imbalanced, AUPRC is the more informative metric~\cite{Saito2015PRMoreInformative}; we additionally report calibration error and bootstrap confidence intervals (Sec.~\ref{ssec:uncal}), though we omit interval values in the main table for space.

\begin{table}[h]
\centering
\caption{Validation performance summary across model families.}
\label{tab:results}
\begin{tabular}{lcc}
\toprule
\textbf{Model} & \textbf{AUROC} & \textbf{AUPRC} \\
\midrule
CVQRC (M=24, V=6) & 0.6871 & 0.4630 \\
ResMix (gate+CV concatenation) & 0.6372 & 0.4165 \\
Random Forest (600 trees) & 0.6134 & 0.3789 \\
GateQRC (q=22, d=5, inj=0.85) & 0.5920 & 0.3649 \\
SVM (RBF, $C=10$) & 0.5727 & 0.3594 \\
\bottomrule
\end{tabular}
\end{table}

A few observations follow immediately. First, even in fully quantum-inspired mode, the CV reservoir family is consistently the strongest. Second, the gate reservoir is more sensitive to hyperparameters (qubit count, depth, injection fraction); the best variant we tried---22 qubits, depth 5, 85\% injected---breaks 0.59 AUROC and 0.36 AUPRC, but some deeper/larger variants hurt, confirming the observations in~\cite{FujiiNakajima2017Harness}. Third, the classical RF remains a strong AUROC baseline, which is useful because it gives us a clear, well-understood reference point when assessing the added value of quantum-inspired reservoirs under different cleaning strategies and subject-wise splits.

\begin{figure}[h]
    \centering
    \includegraphics[width=0.75\linewidth]{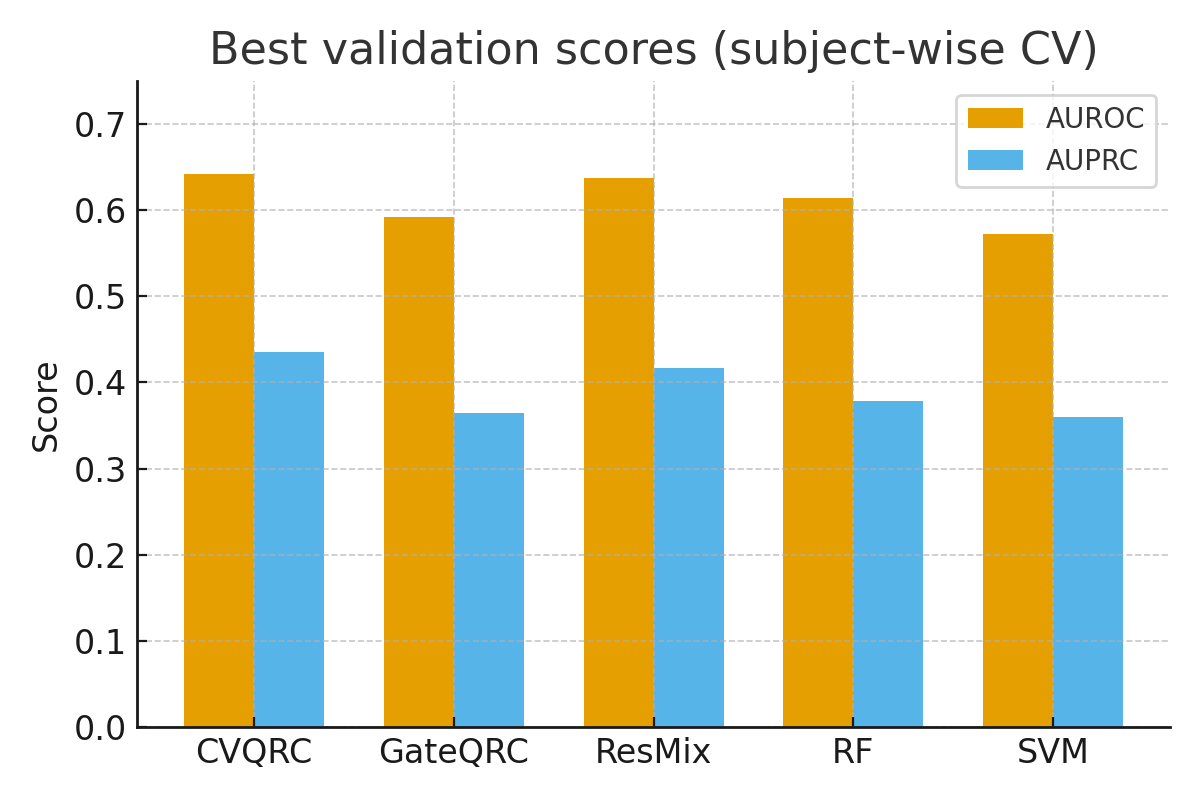}
    \caption{This plot evaluates the performance of all the models, and it can be clearly noted that the quantum-inspired reservoirs outperform the baselines, with CVQRC being the best model}
    \label{results_plot}
\end{figure}

\section{Discussion}
Our study is fully \emph{quantum-inspired} (classical simulation), so we do not claim quantum advantage. Rather, we test whether quantum-shaped reservoir dynamics provide a useful, lightweight sequence embedding with a simple linear readout. Under strict subject-wise validation, the CV reservoir improves AUPRC over RF/SVM baselines (0.463 vs.\ 0.36--0.38) on Helsinki-style feature tables, suggesting the gains are not an artefact of subject leakage; the gate-style variant is more hyperparameter-sensitive. CV models also tended to be better calibrated (ECE $<0.05$ in most folds), which is relevant for alarm thresholding. Future work includes broader multi-seed sensitivity grids, imbalance-aware objectives, and multi-task heads (seizure + artifact/state), and rerunning the same reservoir layouts on higher-fidelity backends when available.

A limitation is that we compare mainly to tabular baselines (RF, RBF--SVM) rather than end-to-end deep sequence models (GRU/TCN) on the same windows; recent neonatal work shows deep models can be strong on raw/quantitative EEG with sufficient data and careful validation~\cite{Kim2025ShortHorizonNeonatal,Hogan2025ScalingNeonatalCNN}. Adding GRU/TCN baselines on the \emph{same} feature-table windows is an important next step to isolate whether improvements come from quantum-shaped dynamics versus generic recurrent dynamics.

Scope is intentionally narrow: neonatal EEG differs from adult/pediatric corpora, so we do not assume identical behavior on datasets such as Bonn or CHB-MIT. Nevertheless, the same encoders can attach to raw-EEG front-ends (e.g., STFT banks) and the window length $L$ provides a direct control knob for testing longer horizons once larger subject-wise cohorts are available~\cite{Kim2025ShortHorizonNeonatal,Wong2023EEGDatasetsReview}.

\subsection{Interpretability}
Interpretability is aided by the linear readout: reshape $\mathbf{w}$ into per-time-step blocks and use $\|\mathbf{w}_t\|_2$ to quantify which parts of the $L$-step window drive the score. Similar grouping over CV modes gives a coarse ``which modes matter'' view.

\section{Conclusion}
We built a fully reproducible, GPU-friendly neonatal EEG pipeline that preserves the \emph{quantum} structure of reservoir computing while running in classical simulation. It supports two quantum-inspired sequence encoders: a gate-style variant that follows the inject--entangle--readout pattern of discrete qubit reservoirs, and a continuous-variable (bosonic) variant based on displacements, interferometric mixing, weak nonlinearity, and time multiplexing. On Helsinki-style (HUH) data the CV/bosonic reservoir was the strongest performer, suggesting that quantum-shaped dynamics can extract clinically relevant structure from short preictal windows. Because the pipeline also reports per-subject AUPRC, calibration curves, and bootstrap confidence intervals, it provides a complete and defensible sequence-encoding evaluation, and its modular design allows the same reservoirs to be rerun on higher-fidelity backends in the future. We intend to release the code and configurations in a public institutional repository; in the interim, materials are available from the authors upon reasonable request.

\bibliographystyle{IEEEbib}
\bibliography{refs}

@article{Temko2011ClinicalNeonatal,
  author  = {Temko, Andriy and Thomas, Elizabeth and Marnane, William and Lightbody, Gordon and Boylan, Geraldine B.},
  title   = {Performance assessment for {EEG}-based neonatal seizure detectors},
  journal = {Clinical Neurophysiology},
  year    = {2011},
  volume  = {122},
  number  = {3},
  pages   = {474--482},
  doi     = {10.1016/j.clinph.2010.06.035}
}

@article{Temko2011EEGAmplitudeFreq,
  author  = {Temko, Andriy and Nadeu, Climent and Marnane, William and Boylan, Geraldine B. and Lightbody, Gordon},
  title   = {{EEG} signal description with spectral-envelope-based speech recognition features for detection of neonatal seizures},
  journal = {IEEE Transactions on Information Technology in Biomedicine},
  year    = {2011},
  volume  = {15},
  number  = {6},
  pages   = {839--847},
  doi     = {10.1109/TITB.2011.2159805}
}

@article{Stevenson2019NeonatalDataset,
  author  = {Stevenson, Nicholas J. and Tapani, Kimmo and Lauronen, Leena and Vanhatalo, Sampsa},
  title   = {A dataset of neonatal {EEG} recordings with seizure annotations},
  journal = {Scientific Data},
  year    = {2019},
  volume  = {6},
  number  = {1},
  pages   = {39},
  doi     = {10.1038/sdata.2019.39}
}

@article{Saito2015PRMoreInformative,
  author  = {Saito, Takaya and Rehmsmeier, Marc},
  title   = {The Precision-Recall Plot Is More Informative than the {ROC} Plot When Evaluating Binary Classifiers on Imbalanced Datasets},
  journal = {PLOS ONE},
  year    = {2015},
  volume  = {10},
  number  = {3},
  pages   = {e0118432},
  doi     = {10.1371/journal.pone.0118432}
}

@inproceedings{Guo2017Calibration,
  author    = {Guo, Chuan and Pleiss, Geoff and Sun, Yu and Weinberger, Kilian Q.},
  title     = {On Calibration of Modern Neural Networks},
  booktitle = {Proceedings of the 34th International Conference on Machine Learning (ICML)},
  year      = {2017},
  pages     = {1321--1330}
}

@article{Efron1979Bootstrap,
  author  = {Efron, Bradley},
  title   = {Bootstrap Methods: Another Look at the Jackknife},
  journal = {The Annals of Statistics},
  year    = {1979},
  volume  = {7},
  number  = {1},
  pages   = {1--26},
  doi     = {10.1214/aos/1176344552}
}

@article{Temko2011SVMNeonatal,
  author  = {Temko, Andriy and Thomas, Elizabeth and Marnane, William and Lightbody, Gordon and Boylan, Geraldine B.},
  title   = {{EEG}-based neonatal seizure detection with support vector machines},
  journal = {Clinical Neurophysiology},
  year    = {2011},
  volume  = {122},
  number  = {3},
  pages   = {464--473},
  doi     = {10.1016/j.clinph.2010.06.034}
}

@inproceedings{Oshea2017NeonatalCNN,
  author    = {O'Shea, Alison and Lightbody, Gordon and Boylan, Geraldine and Temko, Andriy},
  title     = {Neonatal seizure detection using convolutional neural networks},
  booktitle = {2017 IEEE 27th International Workshop on Machine Learning for Signal Processing (MLSP)},
  year      = {2017},
  pages     = {1--6},
  doi       = {10.1109/MLSP.2017.8168193}
}

@article{FujiiNakajima2017Harness,
  author  = {Fujii, Keisuke and Nakajima, Kohei},
  title   = {Harnessing disordered-ensemble quantum dynamics for machine learning},
  journal = {Physical Review Applied},
  year    = {2017},
  volume  = {8},
  number  = {2},
  pages   = {024030},
  doi     = {10.1103/PhysRevApplied.8.024030}
}

@incollection{Nakajima2020QRC,
  author    = {Fujii, Keisuke and Nakajima, Kohei},
  title     = {Quantum Reservoir Computing: A Reservoir Approach Toward Quantum Machine Learning on Near-Term Quantum Devices},
  editor    = {Nakajima, Kohei and Fischer, Ingo},
  booktitle = {Reservoir Computing: Theory, Physical Implementations, and Applications},
  publisher = {Springer Singapore},
  address   = {Singapore},
  year      = {2021},
  pages     = {423--450},
  doi       = {10.1007/978-981-13-1687-6_18}
}

@article{MartinezPena2021IPC,
  author  = {Mart{\'{\i}}nez-Pe{\~{n}}a, Rodrigo and Giorgi, Gian Luca and Nokkala, Johannes and Soriano, Miguel C. and Zambrini, Roberta},
  title   = {Dynamical Phase Transitions in Quantum Reservoir Computing},
  journal = {Physical Review Letters},
  year    = {2021},
  volume  = {127},
  pages   = {100502},
  doi     = {10.1103/PhysRevLett.127.100502}
}

@article{Killoran2019CVQNN,
  author  = {Killoran, Nathan and Bromley, Thomas R. and Arrazola, Juan Miguel and Schuld, Maria and Quesada, Nicol{\'a}s and Lloyd, Seth},
  title   = {Continuous-variable quantum neural networks},
  journal = {Physical Review Research},
  year    = {2019},
  volume  = {1},
  number  = {3},
  pages   = {033063},
  doi     = {10.1103/PhysRevResearch.1.033063}
}

@article{Schuld2015IntroQML,
  author  = {Schuld, Maria and Sinayskiy, Ilya and Petruccione, Francesco},
  title   = {An introduction to quantum machine learning},
  journal = {Contemporary Physics},
  year    = {2015},
  volume  = {56},
  number  = {2},
  pages   = {172--185},
  doi     = {10.1080/00107514.2014.964942}
}

@article{Oshea2018NeonatalCNN,
  author  = {O'Shea, Aoife and Lightbody, Gordon and Boylan, Geraldine B. and Temko, Andriy},
  title   = {Investigating the impact of convolutional neural networks on neonatal seizure detection},
  journal = {Journal of Neural Engineering},
  year    = {2018},
  volume  = {15},
  number  = {5},
  pages   = {056006},
  doi     = {10.1088/1741-2552/aac9b4}
}

@article{Kim2025ShortHorizonNeonatal,
  author  = {Kim, Jonathan and Amorim, Edilberto and Rao, Vikram R. and Glass, Hannah C. and Bernardo, Danilo},
  title   = {Short-horizon neonatal seizure prediction using {EEG}-based deep learning},
  journal = {PLOS Digital Health},
  year    = {2025},
  volume  = {4},
  number  = {7},
  pages   = {e0000890},
  doi     = {10.1371/journal.pdig.0000890}
}

@article{Hogan2025ScalingNeonatalCNN,
  author  = {Hogan, Robert and Mathieson, Sean R. and Luca, Aurel and Ventura, Soraia and Griffin, Sean and Boylan, Geraldine B. and O'Toole, John M.},
  title   = {Scaling convolutional neural networks achieves expert level seizure detection in neonatal {EEG}},
  journal = {npj Digital Medicine},
  year    = {2025},
  volume  = {8},
  number  = {1},
  pages   = {17},
  doi     = {10.1038/s41746-024-01416-x}
}

@article{Wong2023EEGDatasetsReview,
  author  = {Wong, Sheng and Simmons, Anj and Rivera-Villicana, Jessica and Barnett, Scott and Sivathamboo, Shobi and Perucca, Piero and Ge, Zongyuan and Kwan, Patrick and Kuhlmann, Levin and Vasa, Rajesh and Mouzakis, Kon and O'Brien, Terence J.},
  title   = {{EEG} datasets for seizure detection and prediction---A review},
  journal = {Epilepsia Open},
  year    = {2023},
  volume  = {8},
  number  = {2},
  pages   = {252--267},
  doi     = {10.1002/epi4.12704}
}

@article{Nerenberg2025PNR_QRC,
  author  = {Nerenberg, Sam and Neill, Oliver D. and Marcucci, Giulia and Faccio, Daniele},
  title   = {Photon number-resolving quantum reservoir computing},
  journal = {Optica Quantum},
  year    = {2025},
  volume  = {3},
  number  = {2},
  pages   = {201--210},
  doi     = {10.1364/OPTICAQ.553294}
}

@article{Aadhi2025ScalablePhotonicRC,
  author  = {Aadhi, A. and Di Lauro, L. and Fischer, B. and Dmitriev, P. and Alamgir, I. and Mazoukh, C. and Perron, N. and Viktorov, E. A. and Kovalev, A. V. and Eshaghi, A. and Vakili, S. and Chemnitz, M. and Roztocki, P. and Little, B. E. and Chu, S. T. and Moss, D. J. and Morandotti, R.},
  title   = {Scalable photonic reservoir computing for parallel machine learning tasks},
  journal = {Nature Communications},
  year    = {2025},
  doi     = {10.1038/s41467-025-67983-z},
  note    = {Published online 31 Dec 2025}
}

\end{document}